# Prediction of Intrinsic Triferroicity in Two-Dimensional Lattice

Shiying Shen[a], Yandong Ma[a],*, Xilong Xu[a], Baibiao Huang[a], Liangzhi Kou[b], Ying Dai[a],*

[a]School of Physics, State Key Laboratory of Crystal Materials, Shandong University, Shandanan Str. 27, Jinan 250100, China

[b]School of Chemistry, Physics and Mechanical Engineering, Queensland University of Technology, Brisbane, Queensland, 4001, Australia

*Email: yandong.ma@sdu.edu.cn (Y.M.); daiy60@sina.com (Y.D.)

**Abstract**

Intrinsic triferroicity is essential and highly sought for novel device applications, such as high-density multistate data storage. So far, the intrinsic triferroicity has only been discussed in three-dimensional systems. Herein on basis of first-principles, we report the intrinsic triferroicity in two-dimensional lattice. Being exfoliatable from the layered bulk, single-layer $FeO_2H$ is shown to be an intrinsically triferroic semiconductor, presenting antiferromagnetism, ferroelasticity and ferroelectricity simultaneously. Moreover, the directional control of its ferroelectric polarization is achievable by 90° reversible ferroelastic switching. In addition, single-layer $FeO_2H$ is identified to harbor in-plane piezoelectric effect. The unveiled phenomena and mechanism of triferroics in this two-dimensional system not only broaden the scientific and technological impact of triferroics but also enable a wide range of nanodevice applications.

Table of Contents

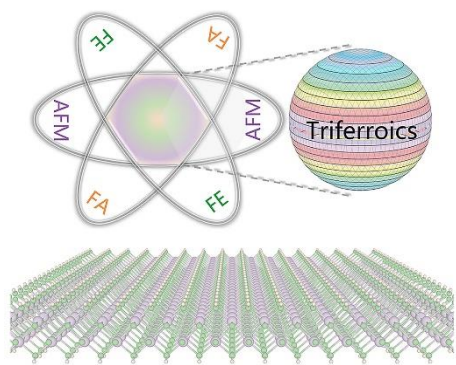

## Introduction

Ferroics is essential to many forms of current and next-generation information technology by virtue of the polarization-switching properties[1-3]. It mainly includes ferromagnets, ferroelectrics and ferroelastics. In each case, crystal symmetry is broken by the coalignment of a particular type of internal degree of freedom, that is, spin orientation for ferromagnets, dipolar displacement for ferroelectrics, and structure deformation for ferroelastics. When two or three ferroic orders coexist in a single-phase material, the multiferroics is achieved[4-8], allowing for a variety of appealing phenomena; for example, the magnetization of a system might be inverted by applying an electric field[6-8]. Beyond ferroic orders, the multiferroics has been broadened to include antiferroic orders, such as antiferromagnets and antiferroelectrics[4,5]. Among different types of multiferroics, undoubtedly, intrinsic triferroics that hold simultaneously three (anti)ferronic orders is most desirable as it possesses unprecedent opportunities for intriguing physics, whose exploitation promises multifunctional and controllable device applications[4-8].

To date, the concept of intrinsic triferroics has only been discussed in three-dimensional (3D) systems[4,5,9]. It is fascinating to note that many fundamental physical phenomena in 3D materials and devices have always found their way to 2D counterparts, such as superconductors[9,10], topological insulators[11-13] and field-effect transistors[14,15]. In general, 2D counterparts have the added advantages of high storage density, low energy consuming, fast device operation and mechanical flexibility. These, combined with technological aspiration of miniaturizing devices, make the achievement of 2D counterparts a highly sought-after target. Actually, many experimental and theoretical studies on 2D ferroics have been reported in the literatures[16-23], and even a few on 2D intrinsic multiferronics that entail two (anti)ferroic orders simultaneously[24-29]. Therefore, an interesting question is whether intrinsic triferroics can be realized in 2D lattice.

In this work, we demonstrate that intrinsic triferroics can indeed be realized in 2D lattice. Based on first-principles calculations, we show that single-layer (SL) $FeO_2H$ is an intrinsically triferroic semiconductor, harboring antiferromagnetism, ferroelasticity and ferroelectricity simultaneously. The physical origin of triferroics has been explained based on the structure and atomic arrangement. This 2D crystal has great thermal and dynamical stabilities as well as easy experimental fabrication from its layered bulk. Furthermore, we predict that such system could demonstrate many distinctive properties, for example, the directional control of ferroelectric polarization by ferroelastic switching and the in-plane piezoelectric effect. These findings will be useful for fundamental research in triferroics and promote technological innovation in nanodevices.

## Results

Magnetism originates from transition-metal atom with partially filled d shells, because the spins of electrons occupying full-filled shells add to zero and do not participate in magnetics. To realize the partially filled d shells, the local structure symmetry, namely coordinated environment, is an essential ingredient. Thus, the combination of transition-metal atom and proper coordinated environment is responsible for the appearance of magnetism. SL $FeO_2H$ is one promising candidate due to the presence of partially filled d shell of Iron. Fig. 1 displays the crystal structure of SL $FeO_2H$, belonging to space group $Pmn2_1$ ($C_{2v}$). The lattice constants are found to be a = 3.79 Å and b = 3.07 Å, with one unit cell containing two Fe, two H and four O atoms. Each Fe atom is coordinated with four oxygen ($O^{2-}$) and two hydroxide ions ($[OH]^-$), forming a distorted octahedral geometry for Fe atom as well as the stoichiometric formula of $[Fe^{3+}] [O]^{2-}[OH]^-$. Given that the valence electronic configuration of an isolated Fe atom is $3d^64s^2$, the half-filled high-spin state of $t_{2g}^3\uparrow e_g^2\uparrow$ is achieved for SL $FeO_2H$, which would give rise to a formal magnetic moment of 5 $\mu_B$ on each Fe atom. As expected, our calculations confirm that the magnetic moment per unit cell (containing two Fe atoms) is 10 $\mu_B$. And the magnetic moment is mainly contributed by Fe atoms, i.e., the magnetic moment on each Fe atom is 4.37 $\mu_B$. This indicates a large spin-polarization in SL $FeO_2H$. We also calculate the low-spin state of SL $FeO_2H$, which is found to be higher in energy than the high-spin ground state by about 0.60 eV/per atom.

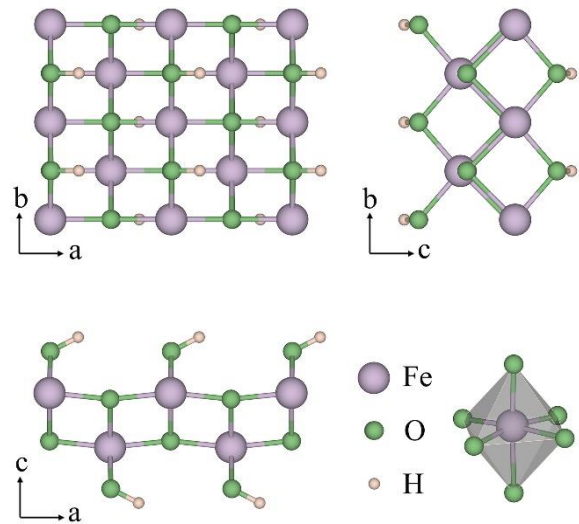

FIG. 1. Crystal structure of SL $FeO_2H$ from top and side views. Insert shows the distorted $FeCl_6$ octahedra.

To probe the preferred magnetic order of SL $FeO_2H$, we consider four configurations, including three antiferromagnetic (AFM) state and one ferromagnetic (FM) state; see Fig. S1a-d. The ground state is found to be AFM-1 shown in Fig. S1b, which is lower than AFM-2, AFM-3 and FM states by 0.11, 0.13 and 0.10 eV/atom, respectively. We also investigate the magnetic ground state of SL $FeO_2H$ by using the PBE functional, and find that AFM-1 remains the ground state, which is lower than that of AFM-2, AFM-3 and FM states by 0.09, 0.04 and 0.06 eV/atom, respectively. Fig. 2a shows the band structure of SL FeO$_2$H. Obviously, SL FeO$_2$H is an AFM semiconductor with an indirect band gap of 2.56 eV. Also, we can see that its conduction band minimum (CBM) is dominated by Fe-3d orbitals, while its valence band maximum (VBM) is mainly derived from O-2p orbitals. We then calculate the magnetic anisotropy energy (MAE) of SL FeO$_2$H. The corresponding results are listed in Table S1, wherein $E_{100}$, $E_{010}$ and $E_{001}$ correspond to the energies with magnetization axis along the (100), (010), and (001) directions. Clearly, the magnetization easy axis is along the y-direction (in-plane), which is lower in energy than that along the x- and z-directions by 167 and 367 μeV per unit cell, respectively. Fig. S2 illustrates the MAE as a function of polar angles in the yz and xy planes. It can be observed that the energies in the yz and xy planes strongly depend on the direction of the magnetization, and the spins in SL FeO$_2$H are favorably aligned along the y direction. These facts imply that SL FeO$_2$H belongs to the category of XY magnets.

The underlying physics of such AFM coupling in SL FeO$_2$H is related to the competition between the direct- and super-exchange interactions. As illustrated in Fig. 2b,c, there are two typical connecting configurations for adjacent octahedrons, i.e. edge-sharing and corner-sharing. In the edge-sharing case, the Fe-O-Fe bond angles are found to be 90° roughly (Fig. 2b). According to the well-known Goodenough-Kanamori-Anderson (GKA) rules[30,31], super-exchange should dominate the interaction between Fe atoms in SL FeO$_2$H, thus leading to FM coupling between nearest-neighboring (NN) Fe atoms (namely, Fe1-Fe2) and between next-nearest-neighboring (NNN) Fe atoms (namely, Fe1-Fe3); see Fig. 2d. Different from the edge-sharing configuration, in corner-sharing configuration, the Fe1-O-Fe4 bond angle is 180° roughly (Fig. 2c). Accordingly, the coupling between next-next-nearest-neighboring (NNNN) Fe atoms (namely, Fe1-Fe4) is AFM; see Fig. 2e. Based on these results, we can easily understand why single-layer FeO$_2$H favors AFM-1 order.

Based on the 2D XY model, the spin Hamiltonian of single-layer FeO$_2$H can be considered as:

$$H = -J_1 \sum_{\langle i,j \rangle} S_i S_j - J_2 \sum_{\langle\langle i,j \rangle\rangle} S_i S_j - J_3 \sum_{\langle\langle\langle i,j \rangle\rangle\rangle} S_i S_j$$

where <i, j>, <<i, j>>, and <<<i, j>>> stand for the NN, NNN, and NNNN sites, respectively. And $J_1$, $J_2$, and $J_3$ are the NN, NNN, and NNNN spin exchanges, respectively; see Fig. S3. Thus, the total energies of different magnetic configurations (scheme in Fig. S1) are:

$$E_{FM} = E_0 - (16J_1 + 8J_2 + 8J_3) \times |S|^2$$
$$E_{AFM1} = E_0 - (8J_2 - 8J_3) \times |S|^2$$
$$E_{AFM2} = E_0 - (-8J_2 + 8J_3) \times |S|^2$$
$$E_{AFM3} = E_0 - (-16J_1 + 8J_2 + 8J_3) \times |S|^2$$

Here, $|S| = \frac{5}{2}$, and the $E_{FM}, E_{AFM1}, E_{AFM2}, E_{AFM3}$ are the total energy of FM, AFM-1, AFM-2, and AFM-3 configurations, respectively. Combining these equations, we find that $J_1$ is FM (1.399 meV), $J_2$ is FM (0.053 meV), and $J_3$ is AFM (−8.959 meV). This is in excellent agreement with the analysis of exchange mechanism.

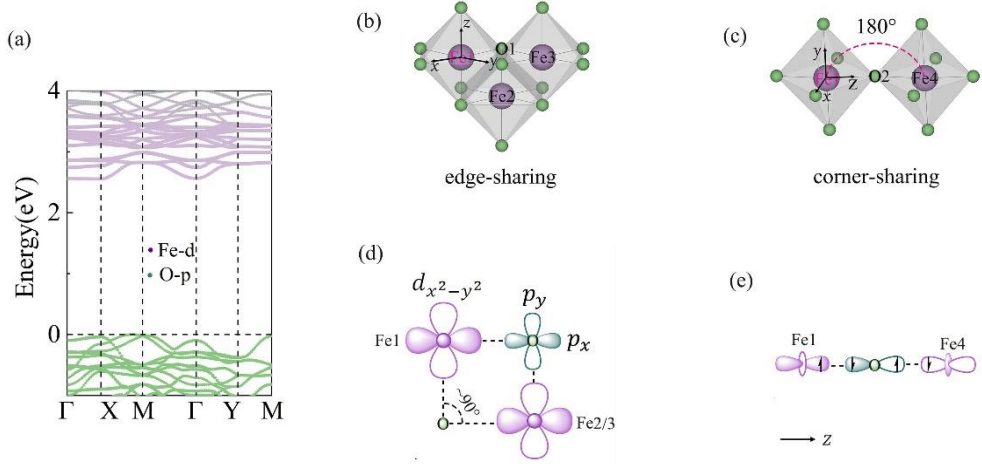

FIG. 2. Magnetism of SL FeO$_2$H. **a** Fat band structure of FeO$_2$H based on HSE06 functional. The Fermi level is shifted to VBM. **b** Edge-sharing configuration for adjacent octahedrons. **c** Corner-sharing configuration for adjacent octahedrons. **d** Schematic diagram of the exchange interaction between Fe1 and Fe2/Fe3 in SL FeO$_2$H. **e** Schematic diagram of exchange interaction between Fe1 and Fe4 in SL FeO$_2$H.

Different from magnetism where local structure symmetry plays an important role, ferroelasticity relates to whole structure symmetry. A necessary condition for realizing ferroelasticity is existence of two or more equally stable orientation variants, which can be switched from one to another in the presence of external stress. Such orientation variants normally originate from the structural phase transition that lowers the symmetry of a prototype phase. As we mentioned above, SL FeO$_2$H belongs to space group *Pmn2$_1$*. This space group can be achieved by reducing the symmetry of space group *Abm$_2$*. And the transformation from space group *Abm$_2$* to space group *Pmn2$_1$* could occur along both x and y directions, forming two different orientations that are perpendicular to each other. In light of these, the appearance of ferroelastic order in SL FeO$_2$H is highly expected.

Fig. 3a shows the two ferroelastic ground states of SL FeO$_2$H, F and F′. In initial variant F, the shorter lattice lies along b axis. By applying uniaxial tensile strain along b axis, the shorter lattice switches to a axis, giving rise to final invariant F′, where the lattice constants are a′ = |b| and b′ = |a|. Obviously, as shown in Fig. 3a, final invariant F′ is similar to initial variant F with a 90° rotation. Inverse transformation between final invariant F′ and initial variant F can also be obtained when imposing uniaxial tensile on final invariant F′ along a axis. The displayed configuration P in Fig. 3a with space group *Abm₂* is supposed to be the paraelastic state of SL FeO$_2$H, with lattice constants a′ = b′ = 3.66 Å. Such paraelastic state would undergo a spontaneous structural transformation to the ground states F and F′.

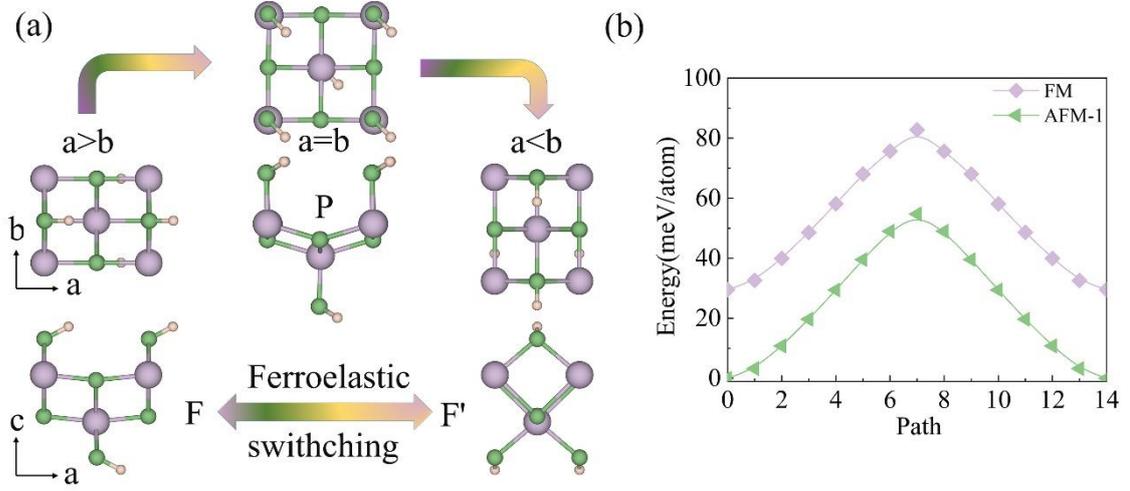

FIG. 3. Ferroelasticity of SL FeO$_2$H. **a** Schematic diagram of ferroelastic switching for SL FeO$_2$H. **b** Minimum energy pathway of ferroelastic switching as a function of step number within NEB for SL FeO$_2$H.

To provide a better physical picture of the ferroelastic behaviors in SL FeO$_2$H, we further investigate the transformation pathway from initial invariant F to final invariant F′ using the nudged elastic band (NEB) method. Fig. 3b shows the minimum energy pathway of ferroelastic switching as a function of step number within NEB. According the symmetry, the energy profile from F to P is identical to that from F′ to P. As shown in Fig. 3b, the magnetic ground state is always AFM-1 during the ferroelastic switching process. The energy barrier for ferroelastic switching in SL FeO$_2$H is estimated to be 54 meV/atom, which is dramatically smaller than that of phosphorene (0.20 eV/atom)[32], borophane (0.10 eV/atom)[33], BP$_5$ (0.32 eV/atom)[34], α-MPI (0.13 eV/atom)[23], VSSe (0.23 eV/atom)[26], and GaTeCl (0.16 eV/atom)[35], and comparable to that of 1T'-WTe$_2$ (73.3 meV/atom)[36]. Such a low barrier renders the high possibility of fast ferroelastic switching in SL FeO$_2$H upon applying external stress. Besides energy barrier, ferroelastic performance also depends on reversible ferroelastic strain, which controls the signal intensity and is defined as $|a/b - 1| \times 100\%$. The obtained reversible ferroelastic strain for SL FeO$_2$H is 23.5%, which is comparable with the values for 1S'-MSSe (4.7%)[22], GeS (17.8%)[32], InOCl (17.7%)[37], Nb$_2$GeTe$_4$ (22.1%) and Nb$_2$SiTe$_4$ (24.4%)[38], suggesting a strong switching signal in SL FeO$_2$H. We therefore demonstrate the promising intrinsic ferroelastics in SL FeO$_2$H. It should be noted that, currently, the 'Curie temperature' for ferroelastic behaviors can only be obtained in experiment and it is still challenging for estimating it theoretically. And of course the temperature will influence the ferroelastic properties, which is not discussed here.

Concerning space group $Pmn2_1$, it has one mirror symmetry only ($M_y$: y→-y). The absence of mirror symmetry $M_x$ for SL FeO$_2$H is attributed to the fact that O-H bond deviates from the c axis by $\theta_{max} = 60.5°$, see Fig. 1. Such deviation is very likely to induce spontaneous polarization in SL FeO$_2$H, and in turn yield ferroelectricity as long as the polarization is switchable. To confirm this speculation, we calculate the total polarization ($P_s$) of SL FeO$_2$H employing the modern theory of polarization based on Berry phase approach[39,40]. The intermediate state, where O-H bond is aligned parallel along the c axis, is taken as the paraelectric state (PE). And the final ferroelectric state is obtained by swirling the angle between O-H bond and c axis from $\theta_{max}$ to -$\theta_{max}$. As shown in Fig. 4a, SL FeO$_2$H exhibits a spontaneous polarization $P_s$ of $0.68 \times 10^{-10}$ Cm$^{-1}$ along a axis. This value is comparable with that of SL As ($0.46 \times 10^{-10}$ Cm$^{-1}$)[41], (CrBr$_3$)$_2$Li ($0.92 \times 10^{-10}$ Cm$^{-1}$)[42] and SnSe ($1.51 \times 10^{-10}$ Cm$^{-1}$)[43], and significantly larger than that of diisopropylammonium bromide ($1.5 \times 10^{-12}$ Cm$^{-1}$)[44] and Hf$_2$VC$_2$F$_2$ ($1.95 \times 10^{-12}$ Cm$^{-1}$)[45].

To further inspect the ferroelectricity, we study the ferroelectric switching process using the NEB method. The energy profiles of ferroelectric switching as a function step number is plotted in Fig. 4b. The switching barrier $E_b$ for SL FeO$_2$H is estimated to be 0.457 eV/unit cell, which is higher than that of BP$_5$ (0.41 eV/unit cell)[34], but much smaller than that of GaTeCl (0.754 eV/unit cell)[35]. Such moderate barrier enables the feasibility of ferroelectric switching in SL FeO$_2$H. And as shown in Fig. S4, the magnetic ground state is always AFM-1 during the ferroelectric switching process.

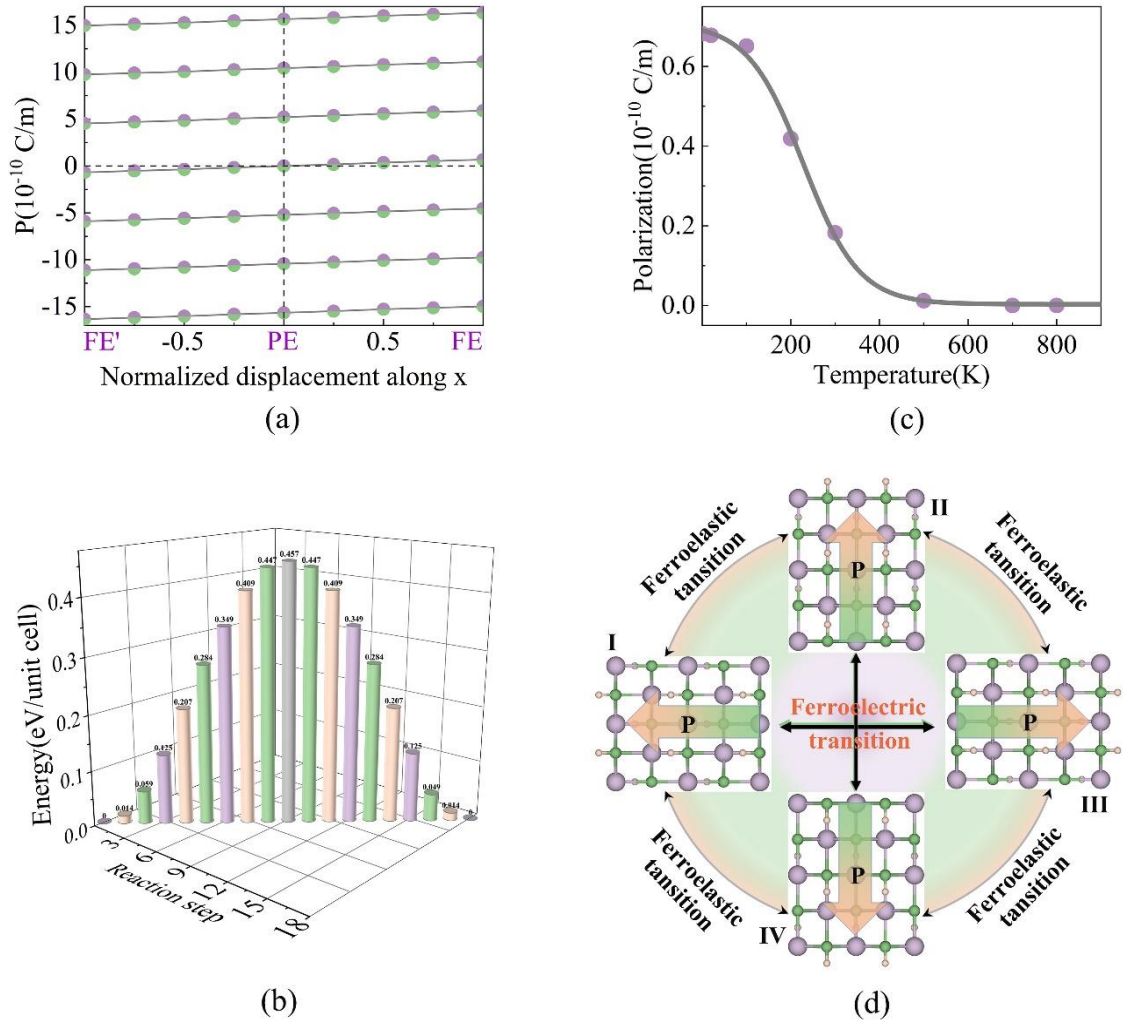

FIG. 4. Ferroelectricity of SL FeO$_2$H. a Total Polarization as a function of normalized displacement for SL FeO$_2$H. b Energy profile of ferroelectric switching as a function of step number within NEB for SL FeO$_2$H. c Polarization as a function of temperature for SL FeO$_2$H obtained from MD simulations. d Schematic diagram of the manipulation of the ferroelectric orders by ferroelastic in SL FeO$_2$H. Fat arrows in d denote the directions of spontaneous polarization.

The ferroelectricity in SL FeO$_2$H can also be described by Landau theory[46]. The potential energy expressed in Landau-Ginzburg-type expansion is given by Bruce[47], which can be considered as the Taylor series of local structural distortions with a certain polarization defined at each cell P$_i$. The first three terms relate to energy contribution from local modes up to sixth order, which are viewed as the Landau energy part, while the last term associates with interaction between the nearest local modes and can be given by mean-field theory within nearest-neighbor approximation. The fitted energy versus polarization is shown in Fig. S5, from which we can see a typical anharmonic double-well energy curve, again confirming the ferroelectricity in SL FeO$_2$H. It should be noted that the existence of ferroelectricity does not guarantee its stability with temperature. Spontaneous polarization would be suppressed when the temperature is above Curie temperature T$_C$. To this end, we investigate the

Curie temperature $T_C$ for ferroelectricity of SL FeO$_2$H. By performing the MD simulations[48], the Curie temperature of SL FeO$_2$H is estimated to be 260 K [Fig. 4c], suggesting its practical applications in 2D memory devices.

**Discussion**

From above, we reveal that SL FeO$_2$H is an intrinsically 2D triferroic semiconductor for the first time. These three ferroic orders could independently encode information in a single bit, meeting the long-pursuit of multifunctional and high-density devices. Concerning using ferroelasticity to encode information, each ferroelastic phase corresponds to one information state[49]. The ferroelastic phase of SL FeO$_2$H has two invariants with different spontaneous strain. Such difference in spontaneous strain renders external stress to couple energetically with the strain state of the system and drive orientation switch, which is analogous to the switching of spontaneous polarization by external electric field in a ferroelectric material[36]. As it is also a major experimental undertaking to measure ferroelastic hysteresis[50]. We wish to point out that such coexistence of ferroic orders can also lead to fascinating physics and applications. For example, it is possible to realize the precise direction-control of ferroelectric polarization P$_s$ in SL FeO$_2$H through ferroelastic switching. As illustrated in Fig. 4d, the lattice structure of SL FeO$_2$H would experience a 90° rotation under the ferroelastic switching, subsequently causing a 90° rotation of polarization P$_s$. According to the symmetry, the direction of polarization P$_s$ in SL FeO$_2$H is switchable among x, y, -x and -y directions through 90° ferroelastic transition. However, it should be noted that the external electric filed cannot trigger ferroelastic switching. Based on this fact, the electrically reading and mechanically writing four-state memory devices are coming up. The four-state information is stored by the polarization P$_s$ pointing to four directions (i.e., x, y, -x and -y) that can be controlled by ferroelectric and ferroelastic transitions. Meanwhile, as the two ferroelastic states relate to the ferroelectric polarization P$_s$ pointing to x/-x and y/-y directions, respectively, these two ferroelastic states can be distinguished electrically. This is attractive as it exploits the best aspects of ferroelectric and ferroelastic data storages.

Another interesting phenomenon in SL FeO$_2$H we wish to address is the piezoelectric effect. In light of its symmetry, the unique piezoelectric coefficients for SL FeO$_2$H are $e_{11}$ and $e_{12}$ are calculated. As listed in Table S3, the value of $e_{11}$ is calculated to be 4.17 × 10$^{-10}$ C/m, which is much larger than that measured for SL MoS$_2$[51]. While for $e_{12}$, the value is estimated to be -0.73 × 10$^{-10}$ C/m. According to the expressions that $d_{11} = (e_{11}C_{22} - e_{12}C_{12}/C_{11}C_{22} - C_{12}^2)$ and $d_{12} = (e_{12}C_{11} - e_{11}C_{12}/C_{11}C_{22} - C_{12}^2)$, the piezoelectric coefficients $d_{11}$ and $d_{12}$ are found to be 2.39 and -1.58 pm/V, respectively. Such in-plane piezoelectric effect in SL FeO$_2$H would make it also potential applicable in sensors and energy conversion devices.

At last, we explore the experimental fabrication and stability of SL FeO$_2$H. Bulk FeO$_2$H has been known since 1978[52]. Fig. S7a shows the crystal structure of bulk FeO$_2$H. It crystallizes in the space group Cmc21, showing a layered structure with two slabs in one-unit cell. To fabricate SL FeO$_2$H from layered bulks, mechanical cleavage and liquid exfoliation are the feasible approaches. Our calculated cleavage energy of SL FeO$_2$H, ~1.13 J/m$^2$, is similar to those of Ca$_2$N (1.08 J/m$^2$)[53], GeP$_3$ (1.14 J/m$^2$)[54] and InP$_3$ (1.32 J/m$^2$)[55], which indicates that the cleavage from the layered FeO$_2$H is accessible. To confirm the stability of SL FeO$_2$H, we investigate its phonon dispersions. As displayed in Fig. S7b, no imaginary phonon mode can be observed, demonstrating that SL FeO$_2$H is dynamically stable. We also perform AIMD simulations to evaluate its thermal stability. The structure of SL FeO$_2$H remains intact and the free energy fluctuates slightly during annealing within 6 ps (see Fig. S8), indicating its thermal stability. Furthermore, we estimate its mechanical stability. The elastic constants of SL FeO$_2$H are calculated to be $C_{11}$ = 162.76 N/m, $C_{22}$ = 104.67 N/m, $C_{12}$ = 41.06 N/m, and $C_{44}$ = 18.24 N/m. These values obey the mechanical stability criteria for a tetragonal monolayer: $C_{11}C_{22} - C_{12}^2 > 0$ and $C_{66} > 0$[56]. The stability analysis ensures that the triferroics predicted in SL FeO$_2$H exhibits superior experimental feasibility.

In summary, using first-principles calculations, we reveal that SL FeO$_2$H harbors antiferromagnetic, ferroelastic and ferroelectric orders simultaneously, being an intrinsically 2D triferroic semiconductor. The physical origins of such triferroics are discussed in detail. We also show that such system can demonstrate many fascinating physics and applications, i.e., the directional control of ferroelectric polarization by ferroelastic switching and the in-plane piezoelectric effect. Moreover, SL FeO$_2$H exhibits superior experimental feasibility as well as great stabilities. Our findings unveil the existence of triferroicity in 2D lattice and offer an ideal platform for studying 2D triferroics.

**Methods**

**First principles calculations**

First-principles calculations are performed within the density functional theory as implemented in the Vienna ab initio simulation package[57,58]. The electron exchange-correlation is described by the generalized gradient approximation (GGA) in the form of Perdew-Burke-Ernzerhof (PBE)[59]. The electron-ion interaction is treated by the projected augmented wave (PAW) method[60]. Cutoff energy is set to be 450 eV. The vacuum space is set as 20 Å. The convergence criteria for energy and force are set to 1 × 10$^{-5}$ eV and 0.02 eV Å$^{-1}$, respectively. The Brillouin zone is sampled using the Monkhorst-Pack scheme[61]: 9 × 9 × 1 for structural optimization and 11 × 11 × 1 for self-consistent calculations. The lattice parameters are optimized in the NM, FM and AFM-1 configurations, which shows negligible difference. The HSE06 functional is adopted for accurately calculating band structures[62]. The phonon spectra is calculated based on density functional perturbation theory using the PHONOPY program[63]. AIMD simulations were performed with a 3 × 3 × 1 supercell for 6 ps with a time step of 1 fs using a NVT ensemble. The ferroelectric polarization is evaluated using the Berry phase method[40].

**Conflicts of interest**

The authors declare no competing financial interest.

**Acknowledgements**

This work is supported by the National Natural Science Foundation of China (No. 11804190), Shandong Provincial Natural Science Foundation of China (Nos. ZR2019QA011 and ZR2019MEM013), Shandong Provincial Key Research and Development Program (Major Scientific and Technological Innovation Project) (No. 2019JZZY010302), Shandong Provincial Key Research and Development Program (No. 2019RKE27004), Qilu Young Scholar Program of Shandong University, and Taishan Scholar Program of Shandong Province.

**Data availability**

The datasets that support the findings of this study are available from the corresponding author on reasonable request.

**References**

1. Von Hippel, A., Breckenridge, R. G., Chesley, F. G. & Tisza, L. High dielectric constant ceramics. *Ind. Eng. Chem.* **38**, 1097 (1946).
2. Matthias, B. T. Ferroelectricity. *Science* **113**, 591 (1951).
3. Aizu, K. Possible species of "ferroelastic" crystals and of simultaneously ferroelectric and ferroelastic crystals. *J. Phys. Soc. Jpn.* **27**, 387 (1969).
4. V. K. Wadhawan, *Introduction to Ferroic Materials* (Gordon and Breach, New York, 2000).
5. Eerenstein, W., Mathur, N. D. & J. F. Scott, Multiferroic and magnetoelectric materials. *Nature* **442**, 759 (2006).
6. Spaldin, N. A. & Fiebig, M. The renaissance of magnetoelectric multiferroics. *Science* **309**, 391 (2005).
7. Hur, N. et al. Electric polarization reversal and memory in a multiferroic material induced by magnetic fields. *Nature* **429**, 392 (2004).
8. Sante, D. D., Stroppa, A., Jain, P. & S. Picozzi, Tuning the ferroelectric polarization in a multiferroic metal–organic framework. *J. Am. Chem. Soc.* **135**, 18126 (2013).

9. Xu, C. et al. Large-area high-quality 2D ultrathin Mo2C superconducting crystals. *Nat. Mater.* **14**, 1135 (2015).

10. Bekaert, J., Petrov, M., Aperis, A., Oppeneer, P. M. & Milošević, M. V. Hydrogen-induced high-temperature superconductivity in two-dimensional materials: The example of hydrogenated monolayer MgB$_2$. *Phys. Rev. Lett.* **123**, 077001 (2019).

11. Yu, R. et al. Quantized anomalous Hall effect in magnetic topological insulators. *Science* **329**, 61 (2010).

12. Kane, C. L. & E. J. Mele, Quantum spin Hall effect in graphene. *Phys. Rev. Lett.* **95**, 226801 (2005).

13. Ma, Y. D., Kou, L. Z., Dai, Y. & Heine, T. Proposed two-dimensional topological insulator in SiTe. *Phys. Rev. B* **94**, 201104(R) (2016).

14. Sarkar, D. et al. A subthermionic tunnel field-effect transistor with an atomically thin channel. *Nature* **526**, 91 (2015).

15. Li, L. K. et al. Black phosphorus field-effect transistors. *Nat. Nanotech.* **9**, 372 (2014).

16. Wu, M. H., Dong, S., Yao, K. L., Liu, J. M. & Zeng, X. C. Ferroelectricity in covalently functionalized two-dimensional materials: integration of high-mobility semiconductors and nonvolatile memory. *Nano Lett.* **16**, 7309 (2016).

17. Luo, W. & Xiang, H. J. Two-Dimensional Phosphorus Oxides as Energy and Information Materials. *Angew. Chem.* **128**, 8717 (2016).

18. Chang, K. et al. Discovery of robust in-plane ferroelectricity in atomic-thick SnTe. *Science* **353**, 274 (2016).

19. Gong, C. et al. Discovery of intrinsic ferromagnetism in two-dimensional van der Waals crystals. *Nature* **546**, 265 (2017).

20. Huang, B. et al. Layer-dependent ferromagnetism in a van der Waals crystal down to the monolayer limit. *Nature* **546**, 270 (2017).

21. Miao, N. H., Xu, B., Zhu, L. G., Zhou, J. & Sun, Z. M. 2D intrinsic ferromagnets from van der Waals antiferromagnets. *J. Am. Chem. Soc.* **140**, 2417 (2018).

22. Ma, Y. D., Kou, L. Z., Huang, B. B., Dai, Y. & Heine, T. Two-dimensional ferroelastic topological insulators in single-layer Janus transition metal dichalcogenides MSSe (M=Wo, W). *Phys. Rev. B* **98**, 085420 (2018).

23. Zhang, T., Ma, Y. D., Yu, L., Huang, B. B. & Dai, Y. Direction-control of anisotropic electronic behaviors via ferroelasticity in two-dimensional α-MPI (M = Zr, Hf). *Mater. Horiz.* **6**, 1930 (2019).

24. Huang, C. X. et al. Prediction of intrinsic ferromagnetic ferroelectricity in a transition-metal halide monolayer. *Phys. Rev. Lett.* **120**, 147601 (2018).

25. Yang, L., Wu, M. H., & Yao, K. L. Transition-metal-doped group-IV monochalcogenides: a combination of two-dimensional triferroics and diluted magnetic semiconductors. *Nanotechnology*

**29**, 215703 (2018).

26. Zhang, C. M., Nie, Y. H., Sanvito, S. & Du, A. J. First-principles prediction of a room-temperature ferromagnetic janus VSSe monolayer with piezoelectricity, ferroelasticity, and large valley polarization. *Nano Lett.* **19**, 1366 (2019).

27. Tang, X. & Kou, L. Z. Two-dimensional ferroics and multiferroics: platforms for new physics and applications. *J. Phys. Chem. Lett.* **10**, 6634 (2019).

28. Wu, M. H. & Zeng, X. C. Intrinsic ferroelasticity and/or multiferroicity in two-dimensional phosphorene and phosphorene analogues. *Nano Lett.* **16**, 3236 (2016).

29. Wang, H. & Qian, X. F. Two-dimensional multiferroics in monolayer group IV Monochalcogenides. *2D Mater.* **4**, 015042 (2017).

30. Kanamori, J. Superexchange interaction and symmetry properties of electron orbitals. *J. Phys. Chem. Solids* **10**, 87 (1959).

31. Geertsma, W. & Khomskii, D. Influence of side groups on 90° superexchange: A modification of the Goodenough-Kanamori-Anderson rules. *Phys. Rev. B* **54**, 3011 (1996).

32. Wu, M. H. & Zeng, X. C. Intrinsic ferroelasticity and/or multiferroicity in two-dimensional phosphorene and phosphorene analogues. *Nano Lett.* **16**, 3236 (2016).

33. Kou, L. Z. et al. Auxetic and ferroelastic borophane: a novel 2D material with negative Possion's ratio and switchable dirac transport channels. *Nano Lett.* **16**, 7910 (2016).

34. Wang, H. D., Li, X. X., Sun, J. Y., Liu, Z. & Yang, J. L. BP$_5$ monolayer with multiferroicity and negative Poisson's ratio: a prediction by global optimization method. *2D Mater.* **4**, 045020 (2017).

35. Zhang, S. H. & Liu, B. G. Controllable robust multiferroic GaTeCl monolayer with colossal 2D ferroelectricity and desirable multifunctions. *Nanoscale* **10**, 5990 (2018).

36. Li, W. B. & Li, J. Ferroelasticity and domain physics in two-dimensional transition metal dichalcogenide monolayers. *Nat. Commun.* **7**, 10843 (2016).

37. Xu, X. L., Ma, Y. D., Huang, B. B. & Dai, Y. Two-dimensional ferroelastic semiconductors in single-layer indium oxygen halide InOY (Y= Cl/Br). *Phys. Chem. Chem. Phys.* **21**, 7440 (2019).

38. Zhang, T. et al. Two-dimensional ferroelastic semiconductors in Nb$_2$SiTe$_4$ and Nb$_2$GeTe$_4$ with promising electronic properties. *J. Phys. Chem. Lett.* **11**, 497 (2020).

39. Resta, R. Theory of polarization of crystalline solids. *Rev. Mod. Phys.* **66**, 899 (1994).

40. King-Smith, R. D. & Vanderbilt, D. Theory of polarization of crystalline solids. *Phys. Rev. B* **47**, 1651 (1993).

41. Xiao, C. C. et al. Elemental Ferroelectricity and Antiferroelectricity in Group-V Monolayer. *Adv. Funct. Mater.* **28**, 1707383 (2018).

42. Huang, C. X. et al. Prediction of Intrinsic Ferromagnetic Ferroelectricity in a Transition-Metal Halide Monolayer. *Phys. Rev. Lett.* **120**, 147601 (2018).

43. Fei, R. X., Kang, W. & Yang, L. Ferroelectricity and Phase Transitions in Monolayer Group-

IV Monochalcogenides. *Phys. Rev. Lett.* **117**, 097601 (2016).

44. Ma, L., Jia, Y. L., Ducharme, S., Wang, J. L. & Zeng, X. C. Diisopropylammonium Bromide Based Two-Dimensional Ferroelectric Monolayer Molecular Crystal with Large In-Plane Spontaneous Polarization. *J. Am. Chem. Soc.* **141**, 1452 (2019).

45. Zhang, J. J. et al. Type-II Multiferroic $Hf_2VC_2F_2$ MXene Monolayer with High Transition Temperature. *J. Am. Chem. Soc.* **140**, 9768 (2018).

46. Stanley, H. E. Phase transitions and critical phenomena.
    (*Clarendon Press: Oxford*, 1971).

47. Bruce, A. D. Structural phase transitions. II. Static critical behavior. *Adv. Phys.* **29**, 111 (1980).

48. Barnett, R. N. & Landman, U. Born-Oppenheimer molecular-dynamics simulations of finite systems: Structure and dynamics of $(H_2O)_2$. *Phys. Rev. B* **48**, 2081 (1993).

49. Salje, E. K. H. Phase Transitions in Ferroelastic and Co-Elastic Crystals.
    (*Cambridge Univ. Press*, 1990).

50. Salje, E. K. H. Ferroelastic materials. *Annu. Rev. Mater. Res.* **42**, 265 (2012).

51. Zhu, H. Y. et al. Observation of piezoelectricity in free-standing monolayer $MoS_2$. *Nat. Nanotech.* **10**, 151 (2015).

52. Christensen, H. & Christensen, A. N. Hydrogen bonds of gamma-FeOOH. *Acta Chem. Scand.* **32a**, 87 (1978).

53. Zhao, S. T., Li, Z. Y. & Yang, J. L. Obtaining two-dimensional electron gas in free space without resorting to electron doping: an electride based design. *J. Am. Chem. Soc.* **136**, 13313 (2014).

54. Jing, Y., Ma, Y. D., Li, Y. F. & Heine, T. $GeP_3$: a small indirect band gap 2D crystal with high carrier mobility and strong interlayer quantum confinement. *Nano Lett.* **17**, 1833 (2017).

55. Miao, N. H., Xu, B., Bristowe, N. C., Zhou, J. & Sun, Z. M. Tunable magnetism and extraordinary sunlight absorbance in indium triphosphide monolayer. *J. Am. Chem. Soc.* **139**, 11125 (2017).

56. Wang, J. H. & Yip, S. Crystal instabilities at finite strain. *Phys. Rev. Lett.* **71**, 4182 (1993).

57. Kresse, G. & Furthmuller, J. Efficient iterative schemes for ab initio total-energy calculations using a plane-wave basis set. *Phys. Rev. B* **54**, 11169 (1996).

58. Kresse, G. & Furthmuller, J. Efficiency of ab-initio total energy calculations for metals and semiconductors using a plane-wave basis set. *Comput. Mater. Sci.* **6**, 15 (1996).

59. Perdew, J. P., Burke, K. & Ernzerhof, M. Generalized gradient approximation made simple. *Phys. Rev. Lett.* **77**, 3865 (1996).

60. Kresse, G. & Joubert, D. From ultrasoft pseudopotentials to the projector augmented-wave method. *Phys. Rev. B* **59**, 1758 (1999).

61. Monkhorst, H. J. & Pack, J. D. Special points for Brillonin-zone integrations. *Phys. Rev. B* **13**, 5188 (1976).

62. Heyd, J. & Scuseria, G. E. Hybrid functionals based on a screened Coulomb potential. *J. Chem.*

*Phys.* **118**, 8207 (2003).

63.     Gonze, X. & Lee, C. Dynamical matrices, Born effective charges, dielectric permittivity tensors, and interatomic force constants from density-functional perturbation theory. *Phys. Rev. B* **55**, 10355 (1997).

**Competing interests**

The authors declare no competing financial interest.